\documentclass[conference]{IEEEtran}
\IEEEoverridecommandlockouts
\usepackage{cite}
\usepackage{amsmath,amssymb,amsfonts}
\usepackage{algorithmic}
\usepackage{graphicx}
\usepackage{textcomp}
\usepackage{xcolor}
\usepackage[all]{nowidow}
\usepackage[caption=false]{subfig}
\usepackage{balance}

\def\BibTeX{{\rm B\kern-.05em{\sc i\kern-.025em b}\kern-.08em
    T\kern-.1667em\lower.7ex\hbox{E}\kern-.125emX}}
\begin{document}

\title{Towards Distributed Coordination for Fog Platforms\\
    \thanks{Funded by the Deutsche Forschungsgemeinschaft (DFG, German Research Foundation) -- 415899119.}
}

\author{\IEEEauthorblockN{Tobias Pfandzelter, Trever Schirmer, David Bermbach}
    \IEEEauthorblockA{\textit{Technische Universit\"at Berlin \& Einstein Center Digital Future}\\
        \textit{Mobile Cloud Computing Research Group} \\
        \{tp,ts,db\}@mcc.tu-berlin.de}
}

\maketitle

\begin{abstract}
    Distributed fog and edge applications communicate over unreliable networks and are subject to high communication delays.
    This makes using existing distributed coordination technologies from cloud applications infeasible, as they are built on the assumption of a highly reliable, low-latency datacenter network to achieve strict consistency with low overheads.
    To help implement configuration and state management for fog platforms and applications, we propose a novel decentralized approach that lets systems specify coordination strategies and membership for different sets of coordination data.
\end{abstract}


\section{Introduction}
\label{sec:introduction}

\begin{figure}
    \centering
    \includegraphics[width=0.95\linewidth]{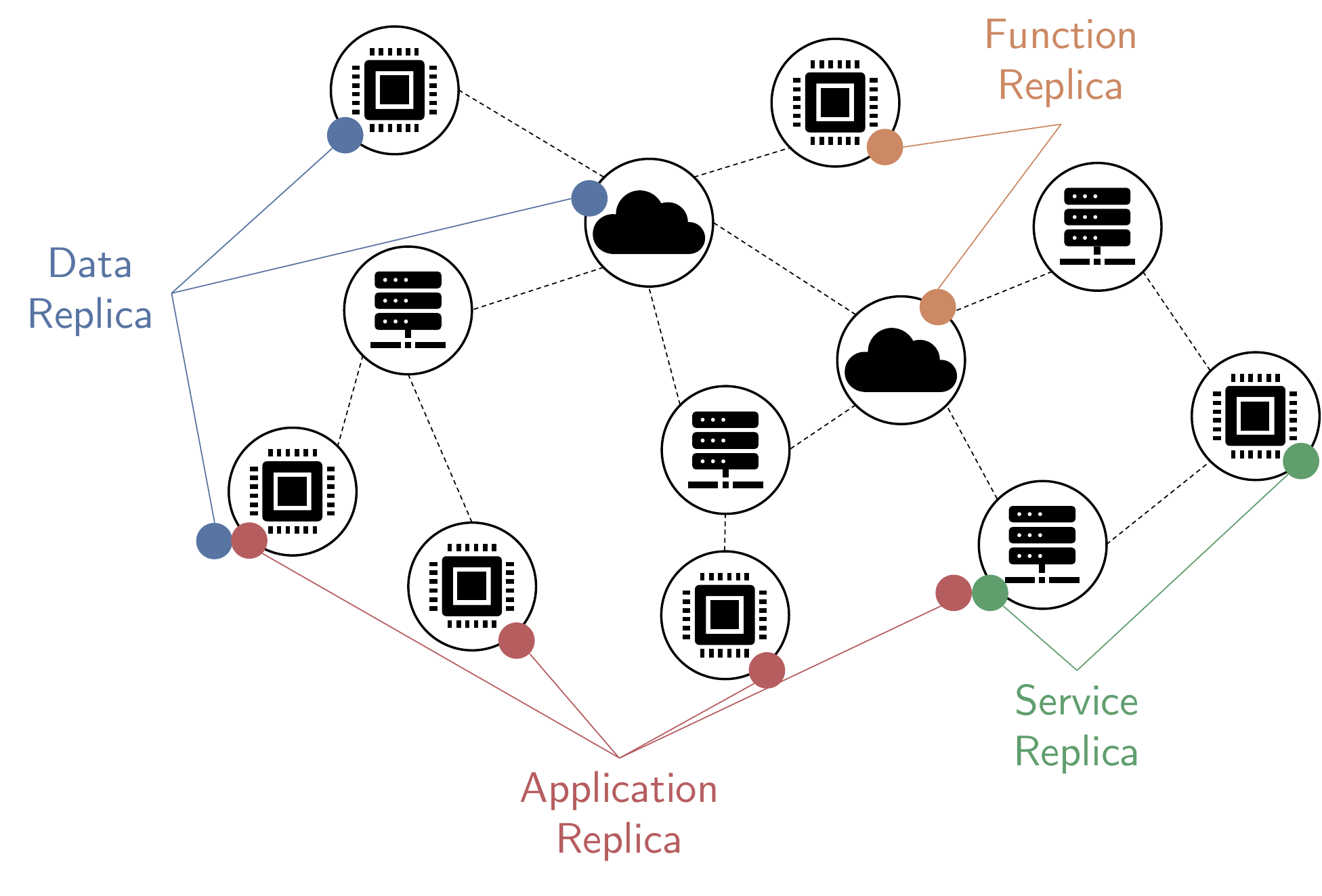}
    \caption{A fog platform manages the replication of application components by abstracting from the underlying geo-distributed and heterogeneous fog infrastructure.}
    \label{fig:platform}
\end{figure}

To leverage fog and edge computing, fog application and data distribution platforms have been proposed that shift the burden of managing the heterogeneous, distributed infrastructure.
As shown in Figure~\ref{fig:platform}, these platforms manage the replication and orchestration of services, e.g., function instances in a FaaS platform~\cite{Cheng2019-zn,Baresi2019-aj,Baresi2019-dt,Bocci2021-gq}, and data, e.g., application state~\cite{Mortazavi2018-wj,Hasenburg2020-yo,Hasenburg2019-oe}.
A key requirement here is the coordination among fog nodes for exchange of configuration and management data, e.g., access control, monitoring data, naming, or routing, not unlike coordination in a distributed cloud system.
Existing distributed application coordination methods and systems, e.g.,~\emph{etcd}\footnote{https://etcd.io/} or \emph{Apache Zookeeper}\footnote{https://zookeeper.apache.org/}, rely on highly available datacenter networks to achieve equally highly available, strictly consistent coordination with low overhead.
In contrast to cloud data centers, fog resources are typically distributed over wide geographical areas and communicate over the Internet instead of dedicated (private) networks, making the existing cloud coordination approaches infeasible:
As a result of CAP~\cite{Gilbert2002-wh} and PACELC~\cite{Abadi2012-us}, the frequent network partitions lead to service disruption, making the platform unavailable.
In addition, the high communication delay between distant fog and edge resources would lead to considerable latency for both centralized and decentralized strictly consistent coordination, either as a result of the long path between a client and central server, or of the path distance between participants.
Finally, fog systems usually comprise thousands of servers in the edge, cloud, and in-between -- orders of magnitudes more than in cloud systems.
These sites are also heterogeneous in their capabilities, from single-board edge computers to clusters of cloud virtual machines~\cite{Bermbach2018-bb}.

Two factors control different quality of service (QoS) aspects of coordination approaches, such as availability and access latency:
(i)~\emph{how} coordination state is exchanged, i.e., synchronously or asynchronously, or which participant may start an update~\cite{wiesmann2000understanding}, and
(ii)~\emph{who} participates in the update exchanges, i.e., the set of nodes that have copies of the coordination data or must reach consensus on an update.
While these factors also apply in cloud applications, their impact in a fog environment is decidedly different, e.g., node membership must also take network distance into account.

The tradeoffs at play here make a one-size-fits-all solution infeasible for fog environments.
The optimal choice depends, among others, on the type of data, data access patterns, the (geo-)distribution of clients accessing the data, and QoS requirements.
In this paper, we thus propose a fog coordination approach that lets developers make these choices per set of coordination data, requiring few changes to the platforms and applications.

\section{Related Work}
\label{sec:relwork}

With an increased research interest in fog computing, numerous fog platforms for application and data management have been proposed that mostly adapt global, strictly consistent coordination from cloud platforms.
\emph{FBase}~\cite{Hasenburg2020-yo,Hasenburg2019-oe} uses a centralized naming service for configuration and access control, similarly to the use of \emph{Chubby}~\cite{Burrows2006-my} in the \emph{GFS}~\cite{Ghemawat2003-gg} and \emph{BigTable}~\cite{Chang2008-xm} data store systems.
In the FBase implementation, this naming service is based on Apache Zookeeper.
\emph{Fogernetes}~\cite{Wobker2018-ud} extends Kubernetes to the fog using a centralized master that handles configuration.
While these approaches allow strictly consistent configuration management without a performance impact for constrained edge devices, the long delay for reads and writes can impact access latency, and any network partition can make the system unavailable.

An alternative approach is a decentralized configuration management.
Eberhardt et al.~\cite{paper_eberhardt_smac} propose using conflict-free replicated data types (CRDTs) to distribute configuration data for Docker containers.
Similarly, Jeffery et al.~\cite{Jeffery2021-hd} propose replacing central configuration in Kubernetes with an eventually consistent, distributed approach.
Using CRDTs, write latency is reduced as conflicts are resolved lazily.
Although decentralized approaches scale well, they also require global synchronization and data exchange, which can be bandwidth-intensive.
Furthermore, the eventual consistency can be problematic for some use-cases, e.g., access control, where a user might expect a permission change to be final.

\section{Distributed Configuration}
\label{sec:architecture}

As we have motivated, in fog platform coordination, different data types require a different tradeoff between latency and consistency.
In contrast to choosing only \emph{either} strict or eventual consistency, our approach lets systems specify how a set of configuration or state data is managed.
Additionally, we propose the notion of coordination memberships, where data is only coordinated within a subset of all fog sites in order to limit update dissemination in the geo-distributed fog environment.
Here, we use the concept of \emph{fog nodes}, a virtual group of machines all running in the same fog location.
Our plan is to implement this approach with a middleware.

\subsection{Coordination Strategy}

Fog platforms require coordination on different types of control data:
Naming data that globally identifies different fog nodes, nodes that replicate a certain service, coordination on data sharding in light of high node churn, or client permissions.
For these different types of control data, it is desirable to be able to specify a coordination strategy to increase consistency where necessary to ensure correct system operation, or to decrease consistency to achieve a lower latency.
We propose between two such strategies, namely:

\subsubsection{Eventual Consistency with CRDTs}

Using CRDTs to manage configuration data allows nodes to write changes without consulting other nodes in the system.
This reduces request latency and keeps the system available in case of network partitions.
While the observed configuration state is guaranteed to converge in the absence of errors and updates, data will often be stale.
This may be achieved with lazy, intermittent synchronization or gossip among nodes.

\subsubsection{Strict Consistency with Consensus}

Reaching consensus on configuration updates incurs a significant communication delay in a geo-distributed system and might even be impossible when the network is partitioned.
This is desirable in cases where the correct operation of the system depends on strictly consistent coordination.
We may achieve strict consistency by using a consensus protocol among nodes or even with a central coordinator (essentially, a primary copy approach~\cite{wiesmann2000understanding}).

\subsection{Coordination Levels}

\begin{figure}
    \centering
    \includegraphics[width=0.95\linewidth]{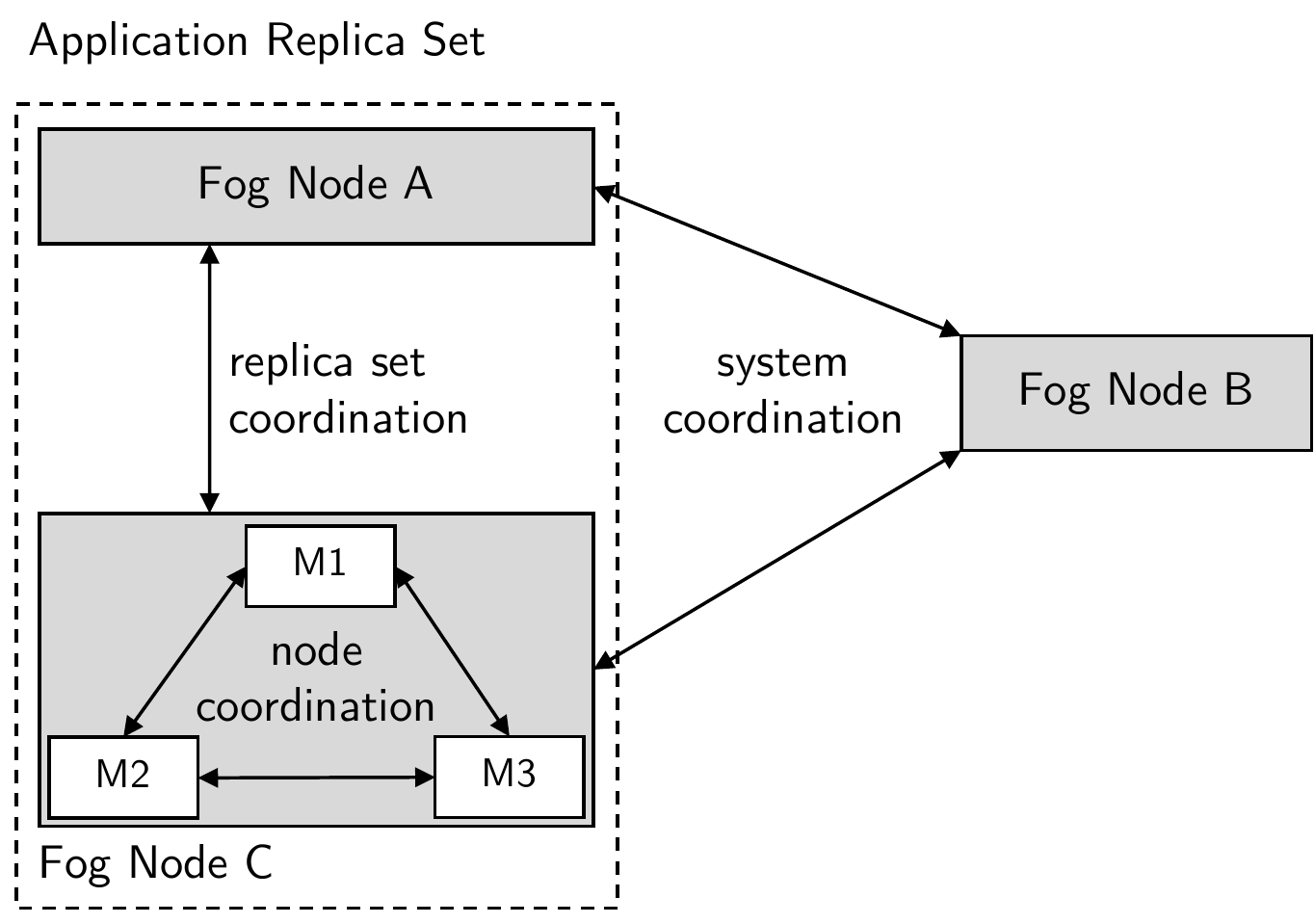}
    \caption{We identify three levels on which fog nodes require coordination: (i) global system coordination among all nodes, required, e.g., for naming; (ii) coordination within a replica set to coordinate, e.g, access control for a service replica; and (iii) node coordination among multiple machines of the same node, e.g., a cluster of datacenter instances.}
    \label{fig:levels}
\end{figure}

Limiting the number of participants for both CRDT message distribution and majority quorums can improve performance and reduce network usage.
We thus introduce different levels of coordination, as shown in Figure~\ref{fig:levels}:
(i) the system level, (ii) the replica set level, and (iii) the node level.
With replica set, we here refer to the group of fog machines jointly managing an application-level replica, e.g., all copies of a data item in the case of a fog storage system, or all instances of a function in case of a FaaS platform.

\subsubsection{System Coordination}
Some data, such as naming data, which must be globally unique and known by all nodes in the system, require coordination among the entire fog system.
As a result, a high communication delay can be expected that imposes a high overhead for consensus with higher write ratios and increases the message dissemination cost for update broadcasts.

\subsubsection{Replica Set Coordination}
Other data is only relevant for members of a replica set, e.g., replica control data.
Such data could be access control lists for a replicated application or membership data for the replica set.
In some cases, where a replica set comprises geographically close nodes, limiting the coordination for this data to a specific set of fog nodes can even reduce the network delay for this coordination group.
Note that the membership of a replica set can change, e.g., when a data replica is migrated to a different location.

\subsubsection{Node Coordination}
A fog node comprises one or more physical machines running in the same location, e.g., a micro-datacenter on the edge.
Within such a fog node, there will often be coordination needs, e.g., when individual machines run different application code.
In this case, the node machines must coordinate in order to appear as one node to other nodes.

\subsection{Architecture}

\begin{figure}
    \centering
    \includegraphics[width=0.95\linewidth]{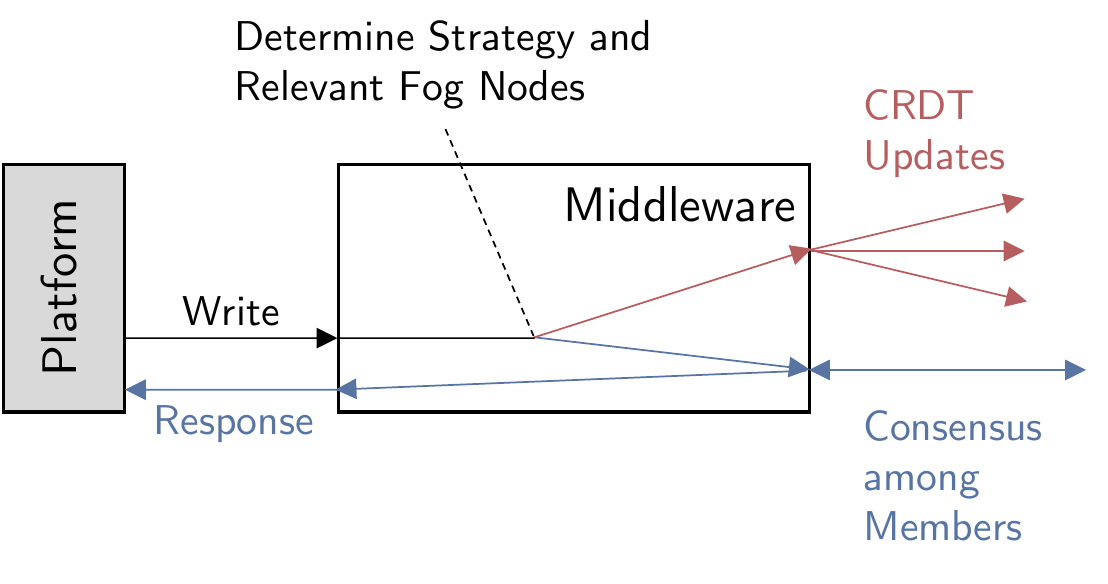}
    \caption{The coordination middleware takes write requests from the application and determines the relevant fog nodes based on the configured level for a data type. Depending on the configured consistency, a CRDT update is sent immediately (red) or consensus is reached with other members (blue).}
    \label{fig:middleware}
\end{figure}

We propose to implement our approach as a coordination middleware that can be used by different fog applications or platforms.
Initially, the application's data types must be configured, i.e., coordination strategies and levels are defined for each coordination data set.
Figure~\ref{fig:middleware} illustrates the update process:
When a write is performed, the middleware first determines the relevant fog nodes for that update depending on membership.
We assume that replica set membership is also stored in this coordination middleware or can be inferred from system configuration data.
For data types that are configured with eventual consistency, the update is propagated to those nodes as a CRDT update in an asynchronous manner.
Where strict consistency is required, a consensus is reached with the relevant nodes before the update is persisted.
The middleware will serve read requests from a local cache, where, e.g., CRDT updates from other nodes are applied.
If required, a read quorum for strictly consistent data items may also be established.
Alternatively, only a subset of the set of member nodes may be chosen, e.g., only the cloud nodes for system coordination, or following a leader-follower approach for node coordination.

\balance
\section{Conclusion \& Future Work}
\label{sec:conclusion}

In this paper, we have proposed a new architecture for distributed coordination in fog platforms.
We have motivated why different types of data require different tradeoffs between consistency and latency.
Further, we have shown different levels of coordination, from system to node configuration and state management, which limits message passing and consensus participants for geo-distributed fog platforms.
As a middleware, an implementation of this approach can be used by different fog and edge platforms without changes to their architecture.

It also opens a number of interesting future research directions:
We plan to evaluate the impact of consistency tuning on both correctness and performance of different fog platforms to find out which kinds of data can benefit from relaxed consistency guarantees.
Of course, the middleware may be extended for other coordination strategies.
Furthermore, fog overlay networks may be used for deduplication of CRDT update messages to limit the network cost of an update broadcast.
Finally, we also plan to explore inferring coordination strategies and levels from data access to allow an integration of the coordination middleware without explicit configuration of data types.

\bibliographystyle{IEEEtran}
\bibliography{bibliography}

\begin{thebibliography}{10}
\providecommand{\url}[1]{#1}
\csname url@samestyle\endcsname
\providecommand{\newblock}{\relax}
\providecommand{\bibinfo}[2]{#2}
\providecommand{\BIBentrySTDinterwordspacing}{\spaceskip=0pt\relax}
\providecommand{\BIBentryALTinterwordstretchfactor}{4}
\providecommand{\BIBentryALTinterwordspacing}{\spaceskip=\fontdimen2\font plus
\BIBentryALTinterwordstretchfactor\fontdimen3\font minus
  \fontdimen4\font\relax}
\providecommand{\BIBforeignlanguage}[2]{{%
\expandafter\ifx\csname l@#1\endcsname\relax
\typeout{** WARNING: IEEEtran.bst: No hyphenation pattern has been}%
\typeout{** loaded for the language `#1'. Using the pattern for}%
\typeout{** the default language instead.}%
\else
\language=\csname l@#1\endcsname
\fi
#2}}
\providecommand{\BIBdecl}{\relax}
\BIBdecl

\bibitem{Cheng2019-zn}
B.~Cheng, J.~Fuerst, G.~Solmaz, and T.~Sanada, ``Fog function: Serverless fog
  computing for data intensive {IoT} services,'' in \emph{Proceedings of the
  2019 {IEEE} International Conference on Services Computing ({SCC} 2019)},
  Jul. 2019, pp. 28--35.

\bibitem{Baresi2019-aj}
L.~Baresi and D.~Filgueira~Mendon{\c c}a, ``Towards a serverless platform for
  edge computing,'' in \emph{Proceedings of the 3rd {IEEE} International
  Conference on Fog Computing ({ICFC} 2019)}, Jun. 2019, pp. 1--10.

\bibitem{Baresi2019-dt}
L.~Baresi, D.~F. Mendon{\c c}a, and G.~Quattrocchi, ``{PAPS}: A framework for
  decentralized self-management at the edge,'' in \emph{Proceedings of the 17th
  International Conference on {Service-Oriented} Computing ({ICSOC} 2019)},
  Oct. 2019, pp. 508--522.

\bibitem{Bocci2021-gq}
A.~Bocci, S.~Forti, G.-L. Ferrari, and A.~Brogi, ``Secure {FaaS} orchestration
  in the fog: how far are we?'' \emph{Computing}, vol. 103, no.~5, pp.
  1025--1056, 2021.

\bibitem{Mortazavi2018-wj}
S.~H. Mortazavi, B.~Balasubramanian, E.~de~Lara, and S.~P. Narayanan,
  ``Pathstore, a data storage layer for the edge,'' in \emph{Proceedings of the
  16th Annual International Conference on Mobile Systems, Applications, and
  Services (MobySys 2018)}, Jun. 2018, p. 519.

\bibitem{Hasenburg2020-yo}
J.~Hasenburg, M.~Grambow, and D.~Bermbach, ``Towards a replication service for
  data-intensive fog applications,'' in \emph{Proceedings of the 35th {ACM}
  Symposium on Applied Computing, Posters Track ({SAC} 2020)}, Mar. 2020, pp.
  267--270.

\bibitem{Hasenburg2019-oe}
------, ``{{FBase}: A Replication Service for {Data-Intensive} Fog
  Applications},'' Tech. Rep., 2019.

\bibitem{Gilbert2002-wh}
S.~Gilbert and N.~Lynch, ``Brewer's conjecture and the feasibility of
  consistent, available, partition-tolerant web services,'' \emph{ACM SIGACT
  News}, vol.~33, no.~2, pp. 51--59, 2002.

\bibitem{Abadi2012-us}
D.~Abadi, ``Consistency tradeoffs in modern distributed database system design:
  {CAP} is only part of the story,'' \emph{Computer}, vol.~45, no.~2, pp.
  37--42, 2012.

\bibitem{Bermbach2018-bb}
D.~Bermbach, F.~Pallas, D.~G. P{\'e}rez, P.~Plebani, M.~Anderson, R.~Kat, and
  S.~Tai, ``A research perspective on fog computing,'' in \emph{Proceedings of
  the 15th International Conference on Service-Oriented Computing ({ICSOC}
  2017) Workshops}, Aug. 2018, pp. 198--210.

\bibitem{wiesmann2000understanding}
F.~Pedone, M.~Wiesmann, A.~Schiper, B.~Kemme, and G.~Alonso, ``Understanding
  replication in databases and distributed systems,'' in \emph{Proceedings of
  the 20th IEEE International Conference on Distributed Computing Systems
  (ICDCS 2000)}, Apr. 2000, pp. 464--474.

\bibitem{Burrows2006-my}
M.~Burrows, ``The chubby lock service for loosely-coupled distributed
  systems,'' in \emph{Proceedings of the 7th Symposium on Operating Systems
  Design and Implementation (OSDI 2006)}, Sep. 2006, pp. 335--350.

\bibitem{Ghemawat2003-gg}
S.~Ghemawat, H.~Gobioff, and S.-T. Leung, ``The google file system,''
  \emph{Operating Systems Review}, vol.~37, no.~5, pp. 29--43, 2003.

\bibitem{Chang2008-xm}
F.~Chang, J.~Dean, S.~Ghemawat, W.~C. Hsieh, D.~A. Wallach, M.~Burrows,
  T.~Chandra, A.~Fikes, and R.~E. Gruber, ``Bigtable: A distributed storage
  system for structured data,'' \emph{ACM Transactions on Computer Systems},
  vol.~26, no.~2, pp. 4:1--4:26, 2008.

\bibitem{Wobker2018-ud}
C.~W{\"o}bker, A.~Seitz, H.~Mueller, and B.~Bruegge, ``Fogernetes: Deployment
  and management of fog computing applications,'' in \emph{Proceedings of the
  2018 {IEEE/IFIP} Network Operations and Management Symposium ({NOMS} 2018)},
  Apr. 2018, pp. 1--7.

\bibitem{paper_eberhardt_smac}
J.~Eberhardt, D.~Ernst, and D.~Bermbach, ``Smac: State management for
  geo-distributed containers,'' in \emph{Proceedings of the 2nd IEEE
  International Workshop on Container Technologies and Container Clouds (WoC
  2016)}, Apr. 2016.

\bibitem{Jeffery2021-hd}
A.~Jeffery, H.~Howard, and R.~Mortier, ``Rearchitecting kubernetes for the
  edge,'' in \emph{Proceedings of the 4th International Workshop on Edge
  Systems, Analytics and Networking (EdgeSys 2021)}, Apr. 2021, pp. 7--12.

\end{thebibliography}

\end{document}